\documentclass[reprint,
    superscriptaddress,
    amsmath,amssymb,prd,aps,twocolumn]{revtex4-2}
\usepackage{booktabs}
\usepackage{hyperref}
\usepackage{graphicx}
\usepackage{dcolumn}
\usepackage{subcaption}
\usepackage{xcolor}
\usepackage{float}
\usepackage{bm}

\begin{document}

\preprint{APS/123-QED}

\title{A Simulation-Based Inference Evaluation of Tension Between MicroBooNE and MiniBooNE Results in a 3+1 Sterile Neutrino Global Fit}


\author{Julia P. Woodward\thanks{First author}}
\email{julia785@mit.edu}
\affiliation{Department of Physics, Massachusetts Institute of Technology, Cambridge, MA 02139, USA}
\author{Joshua Villarreal}
\affiliation{Department of Physics, Massachusetts Institute of Technology, Cambridge, MA 02139, USA}
\author{John M. Hardin}
\affiliation{Department of Physics, Massachusetts Institute of Technology, Cambridge, MA 02139, USA}
\author{Austin Schneider}
\affiliation{
Texas A\&M University, College Station, Texas 77840
}
\author{Janet M. Conrad}
\affiliation{Department of Physics, Massachusetts Institute of Technology, Cambridge, MA 02139, USA}

\date{\today}

\begin{abstract}
Compatibility between different datasets in a global fit is essential for determining whether a chosen model adequately describes the data. In a 3+1 sterile neutrino global fit, long-standing tensions between datasets sensitive to $\nu_e$ appearance and $\nu_e/\nu_\mu$ disappearance indicate a failure of the model to explain the observed data, despite an overall $> 5\sigma$ improvement over the $3\nu$ Standard Model (SM) based on a $\chi^2$ fit. Overall, a global preference for the 3+1 sterile-neutrino hypothesis with significant tension between experiments motivates consideration of more complex models, but these are currently computationally prohibitive to evaluate. This paper is the third in a series aimed at reducing computational cost by developing a Simulation-Based Inference (SBI) framework for global fits. Previous papers focused on rapidly fitting the data sets using frequentist (Feldman--Cousins) and Bayesian approaches, while in this work, we formalize a definition of tension within the SBI framework.

As an example, we perform a full 3+1 fit to the charged-current quasi-elastic neutrino data from the MiniBooNE experiment and the inclusive neutrino data from the MicroBooNE experiment, located on the same beamline. Using experiment-supplied systematics as is, we find these data sets favor 3+1 at $3.6\sigma$ and $1.8\sigma$ respectively, while the tension between the two is $3.3\sigma$, when fit with the SBI procedure. After correcting for normalization differences between data and Monte Carlo in the MicroBooNE $\nu_\mu$ samples, the tension relaxes to $2.2\sigma$, indicating reduced but non-negligible disagreement. The observed tension may reflect both limitations of the 3+1 model in describing the datasets and the presence of systematic effects that impact the experiments differently.  
\end{abstract}

\maketitle
\section{\label{sec:intro}Introduction}
A number of unexplained anomalies have appeared in neutrino oscillation experiments over the past 25 years.  One hypothesis for these anomalies involves the addition of one or more sterile neutrinos to the Standard Model (SM). Sterile neutrinos do not interact via the weak force but can oscillate to active neutrino flavors, disrupting expected 3-flavor oscillation probabilities. 

The simplest extension to the SM, called ``3+1,'' introduces one such sterile neutrino.  Experiments and global-fitters use the 3+1 model because more complex models are highly computationally intensive.  Fits of the 3+1 model to global data using likelihood optimization and applying Wilks' theorem lead to a $>6\sigma$ improvement over the $3\nu$ SM \cite{Hardin:2022muu}. While this quantity measures an overall preference for 3+1, it does not ask if the individual experiments prefer significantly different parameter values, and hence are actually ``in tension''.  Such tension could indicate systematic issues relating to some or all of the experimental results, or that a better model should still be sought.  Thus, both the result of the global fit and internal tests for compatibility must be considered when exploring extensions to the SM, including 3+1.

To quantify tension, the  parameter goodness of fit (PG) statistic proposed by Maltoni and Schwetz \cite{Maltoni:2003cu} is considered the standard for measuring internal model compatibility. The PG statistic is defined as the difference between the global and summed likelihoods of $r$ experimental datasets, each with $P_r$ independent parameters: 
\begin{equation}
    \chi^2_{PG} =\chi^2_{\text{glob}, \text{min}} - \sum_r \chi^2_{r, \text{min}} \label{PGchi}
\end{equation}
which, under certain requirements similar to Wilks' theorem, can be assumed to follow a $\chi^2$ distribution with $\sum_r P_r - P_\text{glob}$ degrees of freedom.

It has been known for nearly a decade that in the 3+1 model, large internal tension between datasets sensitive to anomalous appearance or disappearance indicate that there is no consistent set of parameter values that fits the data well (see, for example, Ref. \cite{Diaz:2019fwt}). The tension is driven by differences in the allowed mass splitting between the first three quasi-degenerate mass eigenstates and the fourth mass eigenstate, $\Delta m_{41}^2$. At $95\%$ confidence, appearance datasets prefer $\Delta m_{41}^2 < 1\,\text{eV}^2$ while disappearance datasets prefer $\Delta m_{41}^2 > 6\,\text{eV}^2$, resulting in a $4.9\sigma$ PG tension, as seen in Fig. 11 of Ref. \cite{Hardin:2022muu}.



To comprehensively explore the nuances of a global fit that is both highly significant and subject to high tension, it is necessary to first address the difficulties of running global fits. The asymptotic approximation given by Wilks' theorem is often used in frequentist likelihood analyses due to its rapid computation of confidence levels (CLs). Unfortunately, this approximation fails to satisfy expected coverage properties of produced CLs in these scenarios, impacting fit interpretability. The solution is to use the trials-based Feldman-Cousins (FC) method \cite{feldman-cousins}.  However, this is computationally infeasible due to the repeated likelihood optimization tasks required.  

A Simulation Based Inference (SBI) approach solves the computational cost problem while maintaining accuracy. SBI is a subfield of machine learning that aims to approximate probabilistic properties of model parameters and simulated data, enabling quick statistical inference \cite{doi:10.1073/pnas.1912789117}. This approach is particularly valuable when the likelihood function is intractable or too complex to optimize efficiently, as is often the case in global fits of sterile neutrino oscillations. In a previous paper, we have demonstrated that Simulation Based Inference (SBI) produces global fit results more than four orders of magnitude faster per grid point than the method of Feldman and Cousins \cite{10.1088/2632-2153/ae040c}. The framework for performing such a fit is outlined in Ref. \cite{10.1088/2632-2153/ae040c}, and code to fit toy models using this method is publicly available \cite{toygithub}.

In this paper, we explore how to apply an SBI-based approach to the question of tension. As in our previous fitting work, we compute the PG tension statistic using SBI to evaluate many simulated trials. This method does not demand that the statistic follow a $\chi^2$ distribution for a given value of degrees of freedom, which may not hold for measurements of sterile neutrino oscillations.

Consistent with our previous studies, we will present the method using a subset of the global fit data.  In Ref. \cite{10.1088/2632-2153/ae040c}, we demonstrated the strategy for frequentist trials-based fitting using data from muon-flavor disappearance experiments.  In Ref. \cite{Villarreal:2025gux}, we demonstrated the complementary Bayesian approach using electron-flavor disappearance experiments.  In this contribution, we will develop methods for evaluating tension between experiments measuring anomalous electron-flavor appearance.  Specifically, we will simultaneously fit the MiniBooNE and MicroBooNE experiments, for good reason: in global fits, the experiment with the largest impact on the internal tension is MiniBooNE, which has a $>4 \sigma$ preference for 3+1, while MicroBooNE finds no evidence for a sterile neutrino, despite lying on the same beamline as MiniBooNE \cite{MiniBooNE:2020pnu, MicroBooNE:2025nll}.  Taken together, the nuances of the joint analysis of these two appearance searches is emblematic of the study of tension within a global fit.

\section{Inputs to the Three-Oscillation-Channels}

In the 3+1 model, there are three relevant oscillation channels: 
 $\nu_\mu \rightarrow \nu_e$ (electron-flavor appearance), $\nu_e \rightarrow \nu_e$ (electron-flavor disappearance), and $\nu_\mu \rightarrow \nu_\mu$ (muon-flavor disappearance).

We model all three oscillation effects for the MiniBoooNE and MicroBooNE experiments, in part because an appearance-only approach can lead to an artificially enlarged allowed parameter space \cite{Brdar:2021ysi}.
This result is our first example of a three-channel fit within the SBI-framework, and represents a necessary step toward a global fit to all short-baseline data.

\subsection{The 3+1 Model}

The oscillation probabilities for each channel are given by:
\begin{eqnarray}
   P(\nu_\mu \rightarrow \nu_\mu) &=& 1-4|U_{\mu 4}|^2 (1- |U_{\mu 4}|^2 ) \sin^2 \Delta_{41} \label{dismu}
\\
    P(\nu_e \rightarrow \nu_e) &=& 1-4|U_{e4}|^2 (1- |U_{e4}|^2 )\sin^2 \Delta_{41} \label{dise}
\\
     P(\nu_\mu \rightarrow \nu_e) &=& 4|U_{\mu 4}|^2|U_{e4}|^2 \sin^2 \Delta_{41} , \label{appe}
\end{eqnarray}
where
\begin{equation}
\Delta_{41}\equiv 1.27 \Delta m^2_{41}L/E. \label{defDelta}
\end{equation}
These probabilities depend on the mass squared splitting between the lightest and heaviest neutrino, $\Delta m^2_{41}=m^2_4-m^2_1$, which sets the frequency of oscillations as a function of neutrino propagation distance, $L$, and neutrino energy, $E$.
The datasets we will fit are sensitive to the peak of the first oscillation maximum if  $\Delta m^2_{41}\sim 1$ eV$^2$. This is large enough to invoke the ``short baseline approximation'', commonly used in global fits, that assumes that the highest mass state is sufficiently heavy to approximate the three lowest mass states as equivalent, hence $\Delta m^2_{41}\approx\Delta m^2_{42}\approx\Delta m^2_{43}$.

The variables $U_{e4}$ and $U_{\mu4}$ represent elements in the $4\times 4$ PMNS mixing matrix extended to include sterile neutrino oscillations.  Although the mixing matrix may have CP violating terms, the 
large $\Delta m^2_{41}$ value for 3+1 leads to negligible sensitivity to CP violation, so the neutrino and antineutrino oscillation probabilities are identical even in the case of appearance. Often, neutrino mixings are expressed in terms of angles, and in the discussion below we will make use of the definition:
\begin{equation}
\sin^2 2\theta_{\mu e} \equiv 4|U_{\mu 4}|^2|U_{e4}|^2. 
\end{equation}

In the remainder of this section, we will introduce the two experiments studied here and describe the generation of the pseudo-data for the SBI-based analysis.  

\subsection{The MiniBooNE Experiment}

MiniBooNE is a short-baseline neutrino detector located at the Fermi National Accelerator Laboratory on the BNB beamline.
To create the BNB beam, 8 GeV protons impinge upon a Beryllium target producing pions that are collimated by a magnetic focusing horn which subsequently decay.
This produces a neutrino beam peaked around 800 MeV that is mainly muon-flavor, with a small electron-flavor content.
 
The MiniBooNE detector lies on the BNB beamline, 514 m from the production target, and consists of a spherical 818 t mineral-oil detector instrumented with 1520 photo-multiplier tubes that detect scintillation and Cherenkov light \cite{MiniBooNE:2020pnu}.

In datasets taken from 2007 to 2022, MiniBooNE consistently reported an anomalous $>4\sigma$ excess of low-energy ($200 \,\text{MeV} < E^{QE}_\nu < 1250\,\text{MeV}$) electron neutrinos and antineutrinos \cite{MiniBooNE:2020pnu, Kamp:2023mjn, Brdar:2021ysi, Kelly:2022uaa}, referred to as the low energy excess (LEE).  Although the dataset contains significant background, the excess cannot be completely described by a combination of the known background contributions. An ``Altarelli Cocktail'' that allows for an unfavorable combination of different systematic effects can weaken but cannot explain the strength of the result \cite{Brdar:2021ysi,Kelly:2022uaa}.

Fits to MiniBooNE data assuming electron-flavor appearance (Eq.~\ref{appe}), but disregarding any disappearance effects, yield allowed regions in 3+1 parameter space consistent with the anomalous $>4\sigma$ result from the Liquid-Scintillator-Neutrino Detector (LSND) in 2001 \cite{LSND:2001aii}.

Ref.~\cite{Kamp:2023mjn} has pointed out that the angular distribution of the LEE points to a contribution from anti-electron-neutrinos. However, the rates required to explain the excess are too high to be attributed to beam-related backgrounds and such an anti-neutrino component cannot be accommodated within a ``vanilla'' 3+1 scenario without CP violation.

\subsection{The MicroBooNE Experiment}

MicroBooNE is able to set the most stringent exclusions on the MiniBooNE allowed region of any experiment to-date \cite{MicroBooNE:2025nll}. The MicroBooNE detector lies along the BNB beamline, upstream of MiniBooNE, 470 m away from the production target. MicroBooNE collected data from 2015 to 2020 with the primary physics goal of investigating the LEE. Its detector consists of a 170 t Liquid Argon Time Projection Chamber, giving it excellent reconstruction capabilities to avoid particle misidentification to which MiniBooNE may be susceptible. The two experiments are thus complementary. It should be noted that argon has a very low antineutrino cross section, so will not be responsive to anti-electron-neutrino signatures, which may be helpful in understanding the MiniBooNE result \cite{Kamp:2023mjn}. 

In this work, we make use of the 2021 data set \cite{MicroBooNE:2021nxr}.  In 2025, MicroBooNE published results using data from a second beamline at Fermilab, the NuMI beam \cite{MicroBooNE:2025nll}. However, the data release for the 2025 publication does not provide the per-event information required for oscillation reweighting, which is particularly important since the long NuMI beamline represents an extended source \cite{MicroBooNE:2025nll, uBprivate}.

The 2021 sample is the all-inclusive sample analyzed using WireCell (WC) reconstruction. We present an analysis of the exclusive sample of charged current quasi-elastic (CCQE) scattering interactions, analyzed using Deep Learning (DL) reconstruction, in Appendix \ref{sec:dlanalysis} for direct comparison with Ref \cite{MiniBooNE:2022emn}. 

The WC dataset is systematics-limited, but has high statistics and is the primary constraint in MicroBooNE analyses. The sample is sensitive to all three oscillation channels, and contains both fully contained (FC) and partially contained (PC) $\nu_e$ and $\nu_\mu$ samples. The $\pi^0$ samples do not have sufficient energy reconstruction information for accurate simulation studies \cite{MiniBooNE:2022emn}.

A combined fit to MicroBooNE and MiniBooNE data using likelihood maximization and assuming Wilks' theorem was published by the MiniBooNE collaboration in Ref. \cite{MiniBooNE:2022emn}. This result, and its associated data release, is used for comparison and discussion herein.

\subsection{Generating Pseudo-data}

SBI relies on a generator of pseudo-experimental data to build inference models. In this work, pseudo-experiments are generated for MiniBooNE and MicroBooNE following covariance-matrix-based uncertainties made available through each experiment's data releases and the methodology described in \cite{MiniBooNE:2022emn}. 

Following the approach of Ref. \cite{MiniBooNE:2022emn}, we include $\nu_\mu$ samples in both MiniBooNE and MicroBooNE datasets to obtain a full constraint on the oscillation parameter space and indirectly account for the correlations between the two experiments.

While the MiniBooNE prediction is determined using the MiniBooNE $\nu_\mu \rightarrow \nu_e$ simulation, no such information is currently available for MicroBooNE from public data releases. Instead, the MicroBooNE $\nu_e$ prediction is obtained using the MiniBooNE BNB simulation with modified baseline to compute a ratio between the nominal $\nu_e$ intrinsic background prediction and the $\nu_e$ prediction with oscillation effects as a function of true neutrino energy. The intrinsic $\nu_e$ prediction obtained from released MicroBooNE $\nu_e$ simulation is multiplied by this ratio and then added to a constant background to obtain the total prediction in the electron flavor channel. The result of this procedure was shown to agree well with the MicroBooNE publications \cite{MiniBooNE:2022emn}.

\par The pseudo-data used in this study were generated from physics parameters distributed uniformly in logarithmic space on a $50 \times 50 \times 50$ parameter grid, with $U_{e4}$ and $U_{\mu 4}$ spanning $[0.001, 1/\sqrt{2}]$, and $\Delta m_{41}^2$ spanning $[0.01, 100]$. We generate $1000$ pseudo-experiments per parameter point. 

\section{\label{sec:Framework}Simulation-Based Inference Framework}

\subsection{Frequentist Fitting Strategy with Simulation-Based Inference}
The procedure for performing a frequentist fit to data using SBI is outlined in detail in Ref.~\cite{10.1088/2632-2153/ae040c}, but summarized here for completeness. Therein, an analyzer trains a neural network classifier on simulated data to infer the likelihood ratio for pseudo-experimental realization $x$ between two physics parameters $\theta$ and $\theta'$ in a procedure called direct amortized neural likelihood ratio estimation (DNRE) \cite{10.1609/aaai.v38i18.30018}. Outputs from the network can conveniently be manipulated to estimate the posterior density  $P(\theta|x)$ using Bayes' Rule, allowing for the maximum \textit{a posteriori}, $\hat \theta_{\text{MAP}} = \textrm{argmax}_\theta P(\theta|x)$, to function as the best-fit point to satisfy the ranking criterion of the Feldman-Cousins method \cite{feldman-cousins}. This method has been demonstrated to meet expected frequentist coverage guarantees, establishing it as a valid statistical procedure to build confidence levels and associated experimental sensitivities \cite{10.1088/2632-2153/ae040c}. In this analysis, we adopt this method with a few amendments. 

\par For one, we train on parameters $\log_{10} \sin^2 2\theta_{\mu e}$, $\log_{10} \Delta m_{41}^2$, and the log-flavor ratio,  $\log_{10} f_r = \log_{10} U_{e4}/U_{\mu 4}$ to account for the three degrees of freedom. There is a one-to-one mapping between $(\log_{10} \sin^2 2\theta_{\mu e}, \log_{10} f_r)$ and $(\log_{10} U_{e4}$, $\log_{10} U_{\mu 4})$. We profile over physically allowed flavor ratios to generate plots in the familiar $(\sin^2 2\theta_{\mu e}, \Delta m_{41}^2)$ space of 3+1 analyses. The second modification is the replacement of the baked-in DNRE posterior estimation with a sequential neural posterior estimator (SNPE-C) \cite{Villarreal:2025gux, 2019arXiv190507488G}. SNPE-C is a normalizing flow which learns the posterior distribution $P(\theta | x)$ over physics parameters $\theta$ for experimental pseudo-data $x$. In previous work, we found that posterior estimation using SNPE-C is both considerably faster than DNRE and better-performing at our MAP estimation tasks \cite{Villarreal:2025gux}. We train SNPE-C on pseudo-data $x$ and parameters $\log U_{e4}$, $\log U_{\mu 4}$, and $\log \Delta m^2_{41}$ to maintain equivalence $\hat \theta_{MAP} = \hat \theta_{MLE}$. Since the psuedo-data were generated uniformly in these parameters, the associated prior distribution $P(\theta)$ is uniform, allowing for proportionality between the likelihood and posterior distributions via Bayes' rule. As a result, the maximum \textit{a posteriori} estimate coincides with the maximum likelihood estimate.

\subsection{Computing PG Tension with Simulation-Based Inference}
The traditional PG statistic is a difference in log-likelihoods, but as Ref.~\cite{10.1088/2632-2153/ae040c} stands, raw likelihood values are inaccessible via the chosen network framework. Instead, we construct $\chi^2_{PG}$ using a difference in log-likelihood ratios:

\begin{equation}
    \label{eq:chisqpg-diff}
    \begin{split}
    \chi^2_{PG} &= \chi^2_{glob, min} - \sum_r \chi^2_{r, min} \\
         &=\chi^2_{glob, min} - \sum_r \chi^2_{r, min} - (\chi^2_{glob, 0} - \sum_r \chi^2_{r, 0}) \\
     &= \Delta \chi^2_{glob} - \sum_r \Delta \chi^2_{r}
    \end{split}
    \end{equation}
where $\chi^2_0$ is twice the negative log-likelihood evaluated at null ($U_{\mu 4} = U_{\mu e} = 0$). The term in parentheses in the second line of Eq.~\ref{eq:chisqpg-diff} is zero because, under the null hypothesis, the sum of the individual contributions should be identical to the $\chi^2$ of the combined fit. For the remainder of this work, we use $\Delta \chi^2$ to refer to the difference in $\chi^2$ between the best fit and the null hypothesis. For SBI tasks, where likelihood ratios are approximated, we denote the approximate statistic as $\Delta\tilde{\chi}^2$ and corresponding PG statistic as $\tilde{\chi}^2_{PG}$. 
\par To calculate the PG tension, we use 50,000 null realizations to compute $\Delta \tilde{\chi}^2$ values for both the individual experiments and the global fit. Evaluating this number of trials using traditional $\chi^2$ minimization methods would be infeasible. For each set of $\Delta \tilde{\chi}^2$ values, we assign empirical $p$-values based on rank within the simulated distribution, and take the negative logarithm to define a consistent test statistic across different networks, reducing network variability between fits. We can then perform the necessary subtraction to yield $\tilde{\chi}^2_{PG}$. The calibration of $\tilde \chi^2_{PG}$ is discussed in Appendix \ref{sec:calibration}.

\subsection{\label{sec:fits}ML Penalization for Unmodeled Systematic Effects}

In Ref.~\cite{10.1088/2632-2153/ae040c}, we have presented good agreement between most SBI-generated fits and conventional Feldman Cousins-based fits.   We do not expect perfect agreement for a comparison between a Wilks'-based $\chi^2$ fit and an SBI-generated frequentist fit, however those disagreements are modest.
Substantial disagreement between the SBI-generated results and Wilks' results can arise when experiments suffer systematic effects beyond those encoded by the experiments' released covariance matrices. If an unmodeled systematic effect is present in data, but not in the pseudo-experiments used to train the model, a neural network provides test statistic values which inevitably result in more cautious fit results.  

We illustrated this behavior on Anscombe's Quartet of datasets in Ref. \cite{10.1088/2632-2153/ae040c}, and then extended the discussion to oscillation data for SciBooNE and MiniBooNE. There is a systematic downward trend in a combined fit to SciBooNE and MiniBooNE $\bar{\nu}_\mu$ disappearance data that results in the SBI-generated exclusion being weaker compared to the reported exclusion. Such behavior can be viewed as an asset over a $\chi^2$-based approach, where unmodeled systematic effects can sneak through. 

In the cases at hand, both MiniBooNE and MicroBooNE employ covariance matrices and represent systematic uncertainties as $\pm 1 \sigma$ uncertainty bands on their plots. However, the underlying shapes that produce the band are unpublished, and many different shapes of systematic variation can produce the same $\pm 1 \sigma$ band. Observed data within the band may still contain effects that are not encoded in the covariance matrix. If such effects are present, they are expected to manifest as weakened SBI fits, as in the cases discussed above. 

\section{\label{sec:results}Fits Using Simulation-Based Inference}

In this section, we present SBI-generated fits to the MiniBooNE and MicroBooNE datasets. We discuss evidence that both sets of data suffer from unmodeled systematic effects. 
We then present the joint fit results and summarize the fit significances.

\subsection{\label{sec:MBfits}Fit to the MiniBooNE Data Set}

\begin{figure}[h]
    \centering
    \begin{subfigure}{0.85\columnwidth}
    \centering \includegraphics[width=\linewidth]{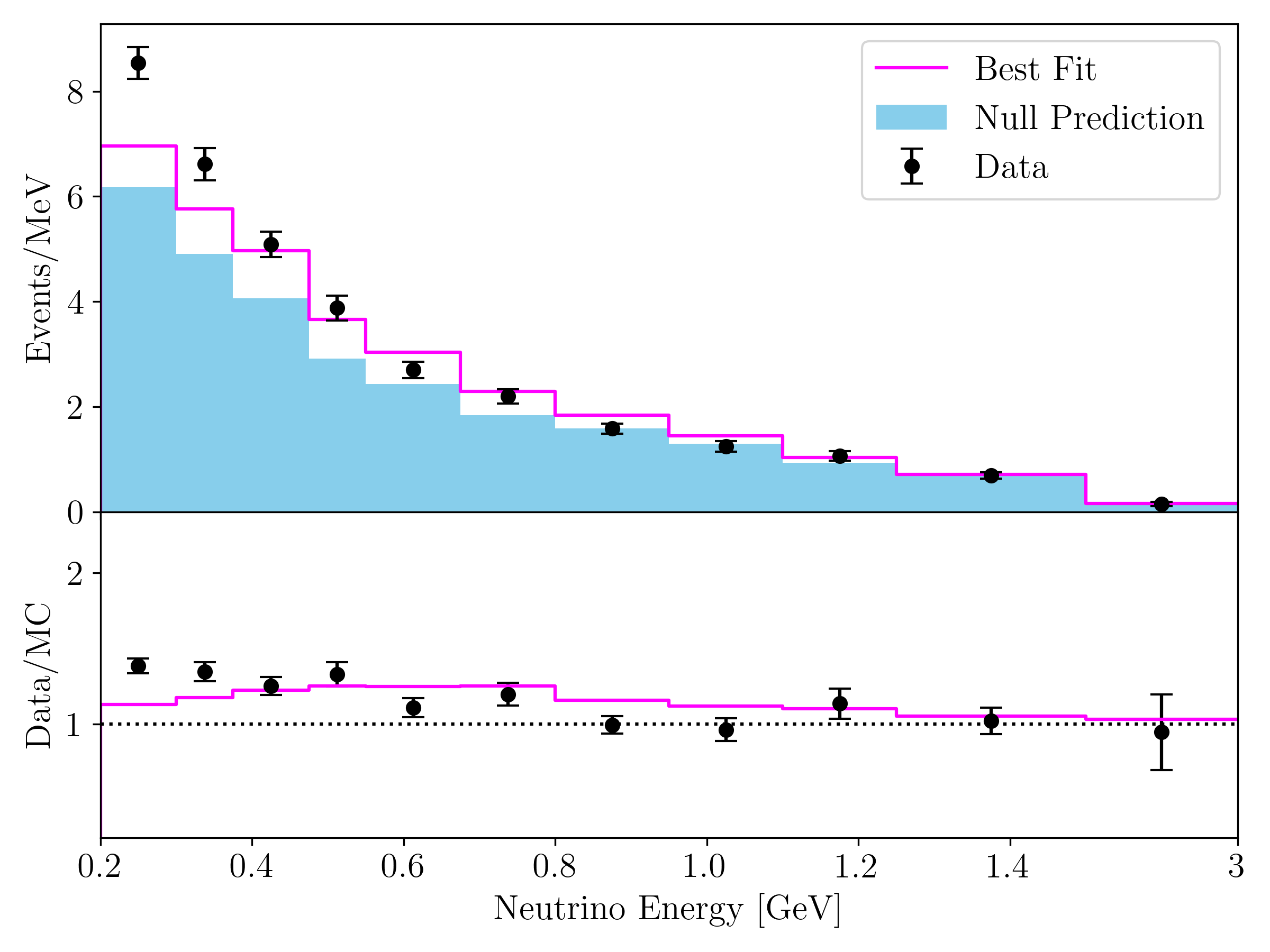}
    \label{fig:datamB}
    \end{subfigure}
    \begin{subfigure}{0.90\columnwidth}
        \centering
        \includegraphics[width=\linewidth]{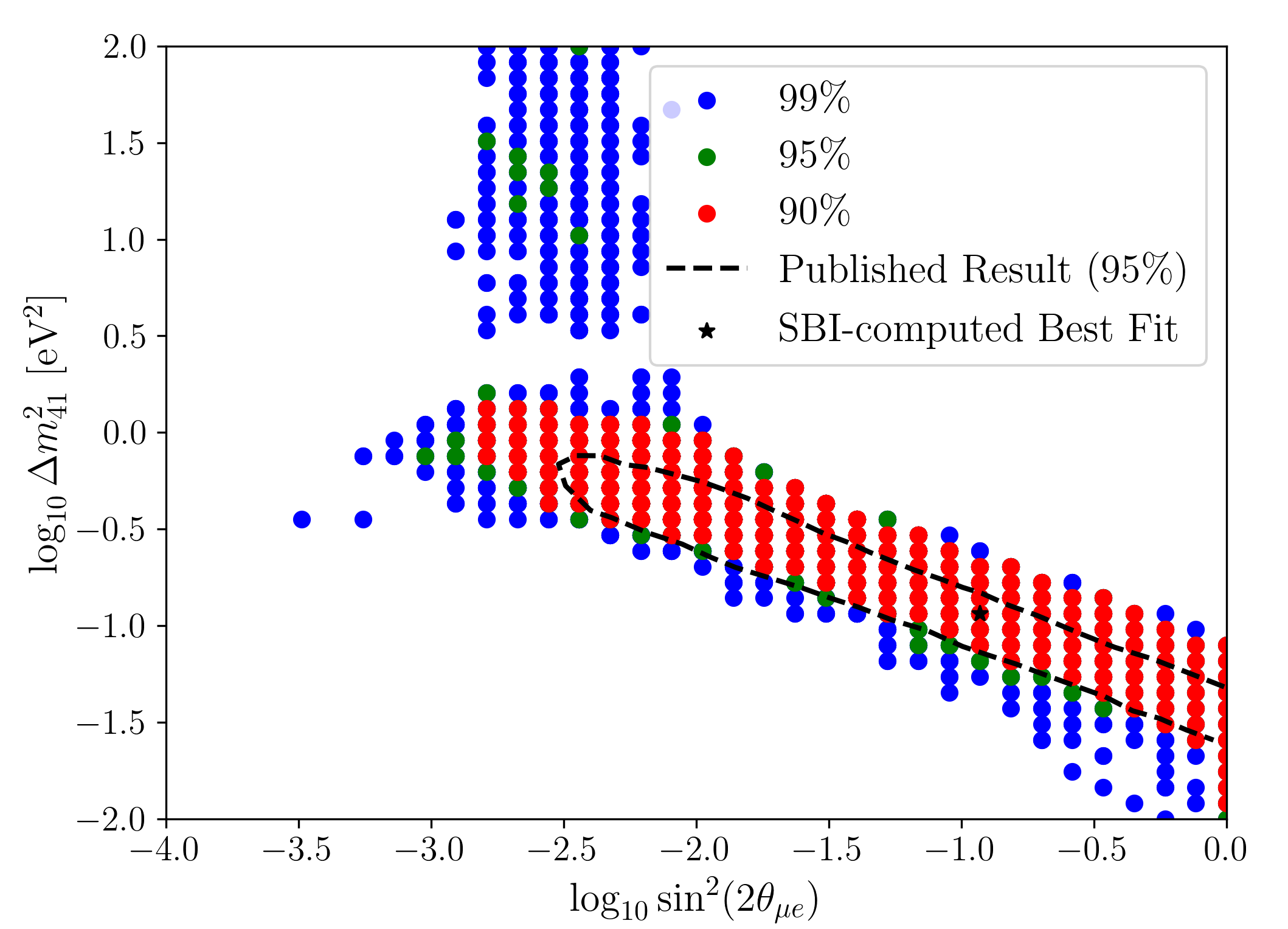}
        \label{fig:mB_fit}
    \end{subfigure}
    
    \caption{(Top) MiniBooNE $\nu_e + \bar{\nu}_e$ data and data/MC ratio distributions, with the Standard Model prediction shown in blue and the SBI-computed best fit 3+1 prediction in pink. Statistical error bars are taken from \cite{MiniBooNE:2022emn}. (Bottom) SBI-computed MiniBooNE 3+1 fit results with the Wilks'-based $95\%$ CL from \cite{MiniBooNE:2022emn} overlaid for comparison.}
    \label{fig:mB}
\end{figure}

In Figure \ref{fig:mB}, we compare the published MiniBooNE Wilks'-based full 3+1 fit results \cite{MiniBooNE:2022emn} to the SBI-computed frequentist confidence levels.  Of the three channels fits, we use $\nu_\mu\rightarrow\nu_e$ appearance as our example. The SBI-computed best fit is $|U_{e4}|^2 = 0.223, |U_{\mu4}|^2 = 0.131$, and $\Delta m_{41}^2 = 0.115$ eV$^2$.  We observe qualitative topological agreement with the published result, but the allowed regions at $90\%$ and $95\%$ CL are wider.  Equivalently, the SBI-based fit for MiniBooNE sets slightly weaker exclusions at high mixing angles than the Wilks'-based fit.  

It is likely that the expanded allowed region is indicative of an unmodeled systematic effect that distorts the data.   As noted above, the MiniBooNE collaboration has already flagged a disagreement between the Wilks'-based 3+1 best fit and the data at low energy, despite a good $p$-value of 12.3\% for 3+1 compared to data, when accounting for statistical and systematic uncertainty \cite{MiniBooNE:2020pnu}.

Similar to the Wilks'-based fit, the SBI-generated best fit does not rise rapidly enough at low energies to fully describe the observed excess (see Fig.~\ref{fig:mB}, which shows the MiniBooNE data with statistical errors only). Such a trend is not present in the pseudo-data used to train the neural network classifier.
Due to this rise at low energy, we conclude that while the MiniBooNE result disfavors null at $3.6 \sigma$ (see Table \ref{tab:significances}), the 3+1 model cannot fully account for the low energy excess.

\subsection{\label{sec:uBfits}Fit to the MicroBooNE Data Sets}
\begin{figure}[h]
    \centering
    \begin{subfigure}{0.85\columnwidth}
    \centering \includegraphics[width=\linewidth]{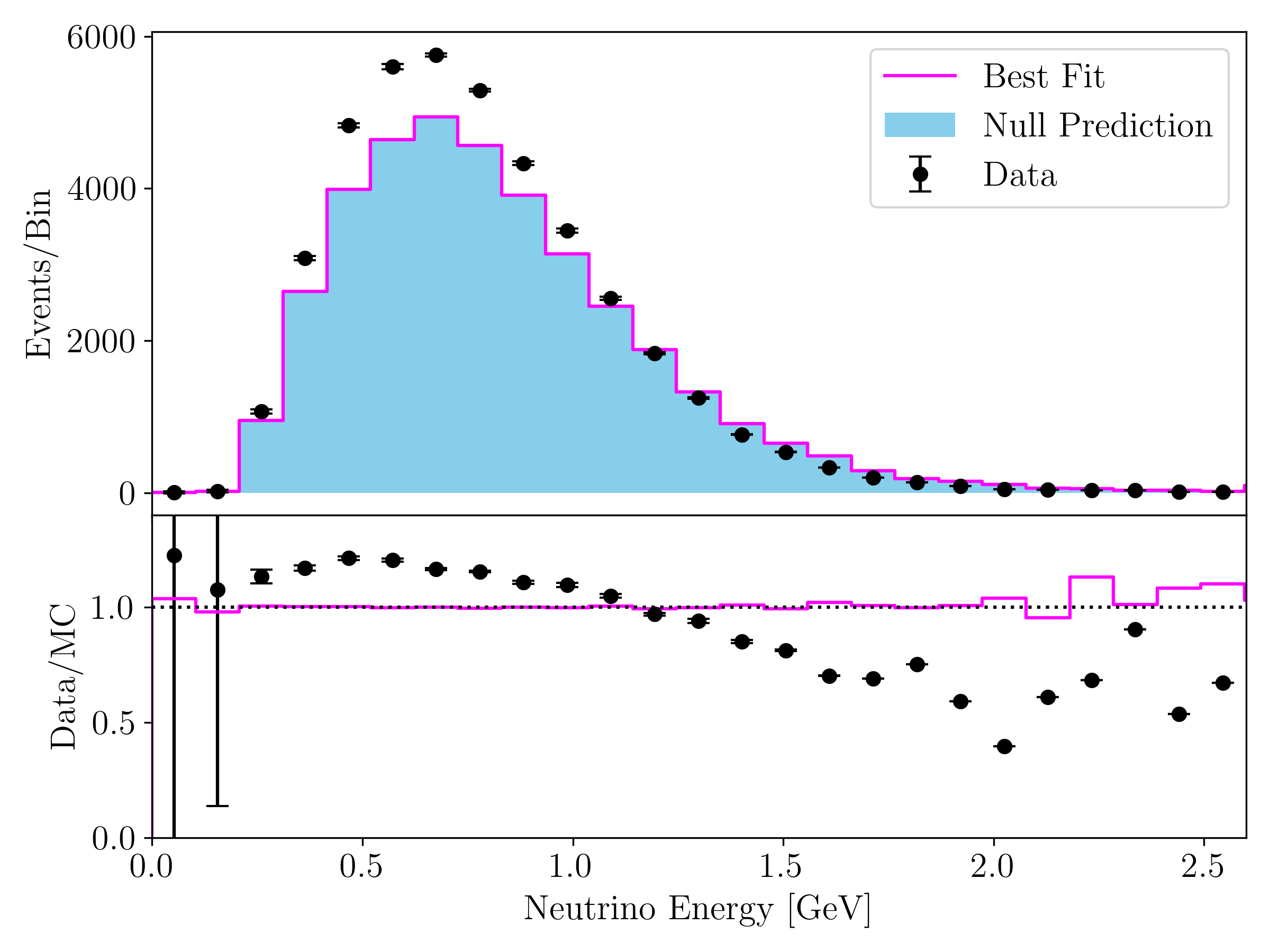}
    \label{fig:dataWCnumuFC}
    \end{subfigure}
    \begin{subfigure}{0.85\columnwidth}
        \centering
        \includegraphics[width=\linewidth]{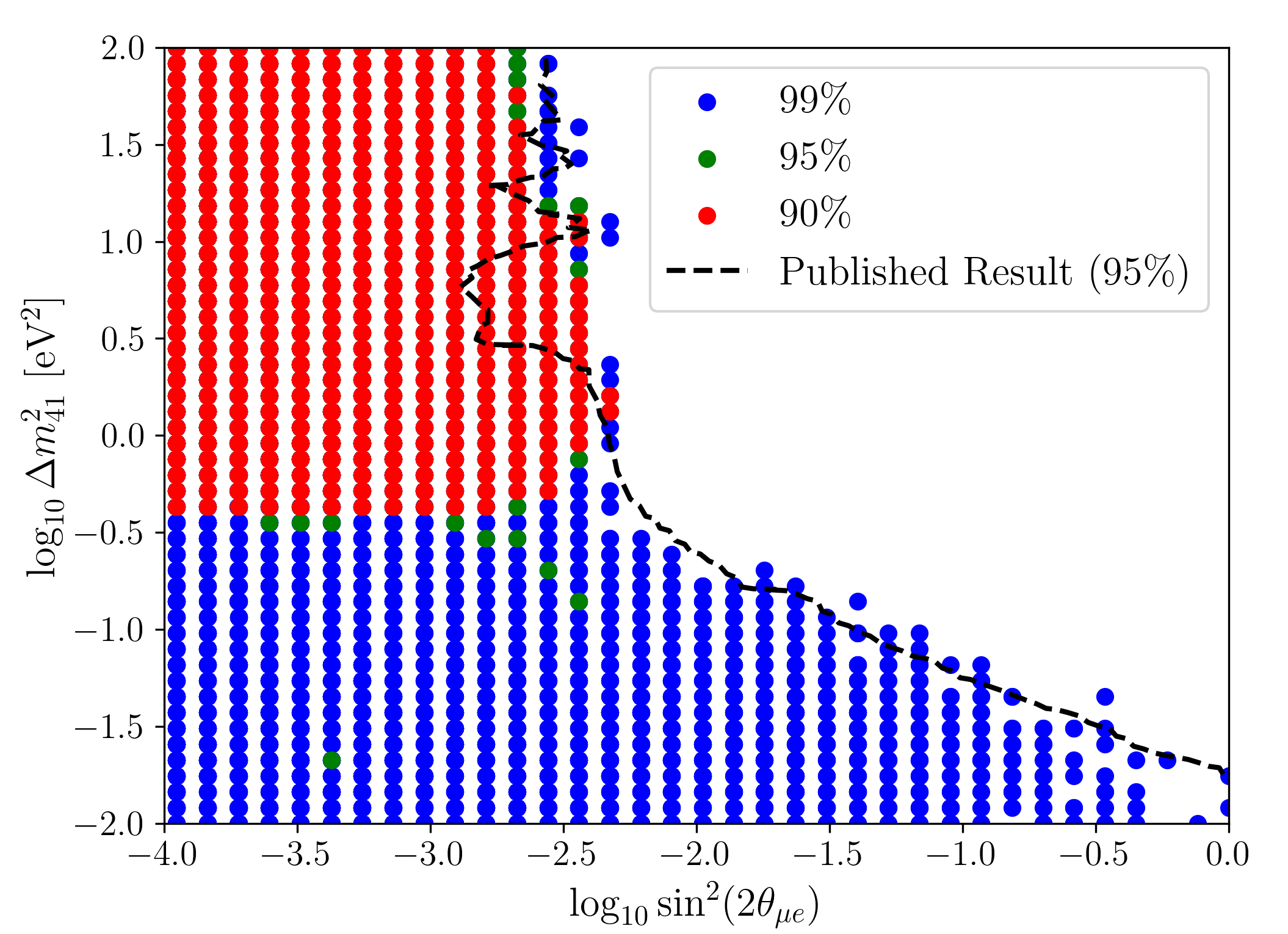}
        \label{fig:WC_fit}
    \end{subfigure}
    
    \caption{Same as Fig. \ref{fig:mB} but for the MicroBooNE $\nu_\mu$ FC dataset, without any additional nuisance parameters added to the fit.}
    \label{fig:WC}
\end{figure}

In the case of MicroBooNE, there is a substantial excess of $\nu_\mu$ events below 1 GeV (see Fig.~\ref{fig:WC}) \cite{MicroBooNE:2021nxr}. In the full 3+1 fit that includes $\nu_\mu$ disappearance, this can lead to an overly strong $\nu_\mu$ disappearance exclusion if the excess is not from a well-described systematic in the covariance matrix. This discrepancy will also affect the $\nu_e$ appearance and disappearance channels, since the high statistics $\nu_\mu$ events are used to predict the $\nu_e$ background rate.   To the level that the excess is relatively flat between 0.25 and 1 GeV, one can think of the discrepancy as setting the overall normalization, pulling up the $\nu_e$ prediction in the fit.

We observe that the SBI-generated MicroBooNE result prefers the 3+1 hypothesis at $1.8\sigma$ (see Tab. \ref{tab:significances}). While this is not significant enough to represent an anomaly, it indicates slight disagreement with the Standard Model. This behavior is illustrated in Fig. \ref{fig:WC}, which shows the SBI-generated fit for the $\nu_\mu \rightarrow \nu_e$ channel. The $90\%$ and $95\%$ CL regions form horizontal bands at low mixing angle, with a sharp transition at  $\Delta m^2 \approx 0.3$ eV$^2$. However, in a typical exclusion, we would expect the fit to be featureless in a region where the experiment does not have sensitivity.  
In fact, a similar but fainter effect was seen in the Wilks'-based fit of Ref.~\cite{MiniBooNE:2022emn}. The SBI-generated fit indicates a systematic issue with the MicroBooNE data that affects the result as a function of $\Delta m^2$, the parameter most sensitive to normalization.

\begin{figure}[h]
    \centering
    \includegraphics[width=0.85\linewidth]{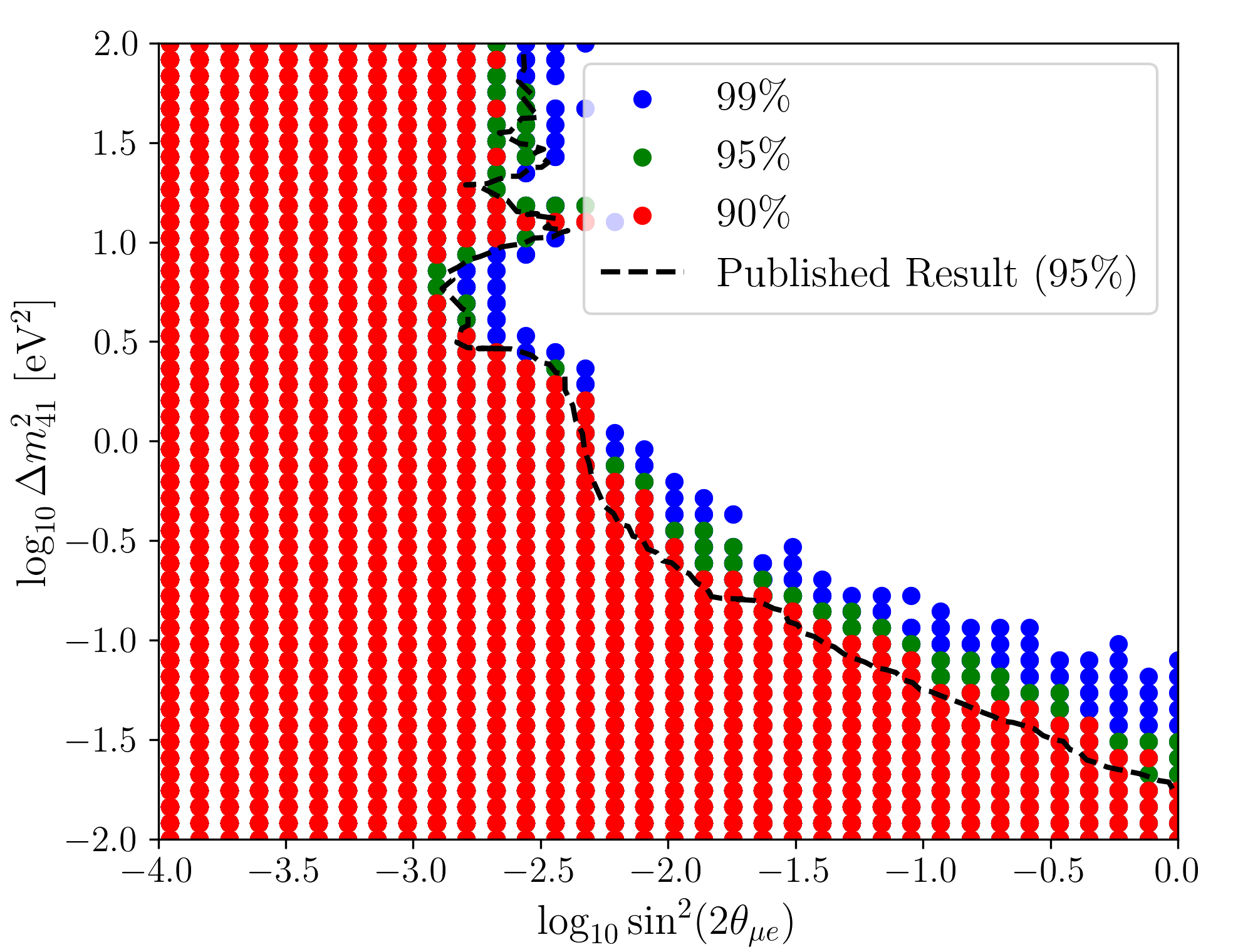}
    
    \caption{SBI-computed fit results for the MicroBooNE dataset with additional nuisance parameters added to the fit in the $\nu_\mu$ samples. We overlay the Wilks'-based $95\%$ CL from \cite{MiniBooNE:2022emn} for comparison.}
    \label{fig:WC_nuisfit}
\end{figure}

We can examine whether this behavior arises from the discrepancy between data and Monte-Carlo (MC) prediction in the $\nu_\mu$ datasets. Refs. \cite{Hardin:2022muu} and \cite{MiniBooNE:2022emn}  have accounted for mismatches in $\nu_\mu$ data by adding nuisance parameters that allow for a relative normalization uncertainty between the $\nu_\mu$ and $\nu_e$ data. If we  adopt this method here, we obtain Fig. \ref{fig:WC_nuisfit}, which shows reasonable agreement with the published results.  Importantly, the confidence level bands now follow the shape of the published contours, as expected.  With these additional nuisance parameters, an SBI-based fit to the MicroBooNE data set finds a $p$-value of $>0.8$, indicating strong agreement with null. 

We conclude that the features present at low mixing angles in Fig.~\ref{fig:WC}, as well as the $1.8\sigma$ preference for 3+1 discussed above, arise from correlated behavior absent from the covariance matrix and thus the network training data. 

\subsection{\label{sec:jointfits}Joint Fits to MiniBooNE and MicroBooNE Data Sets}

\begin{figure}[h]
    \centering

    \includegraphics[width=0.85\linewidth]{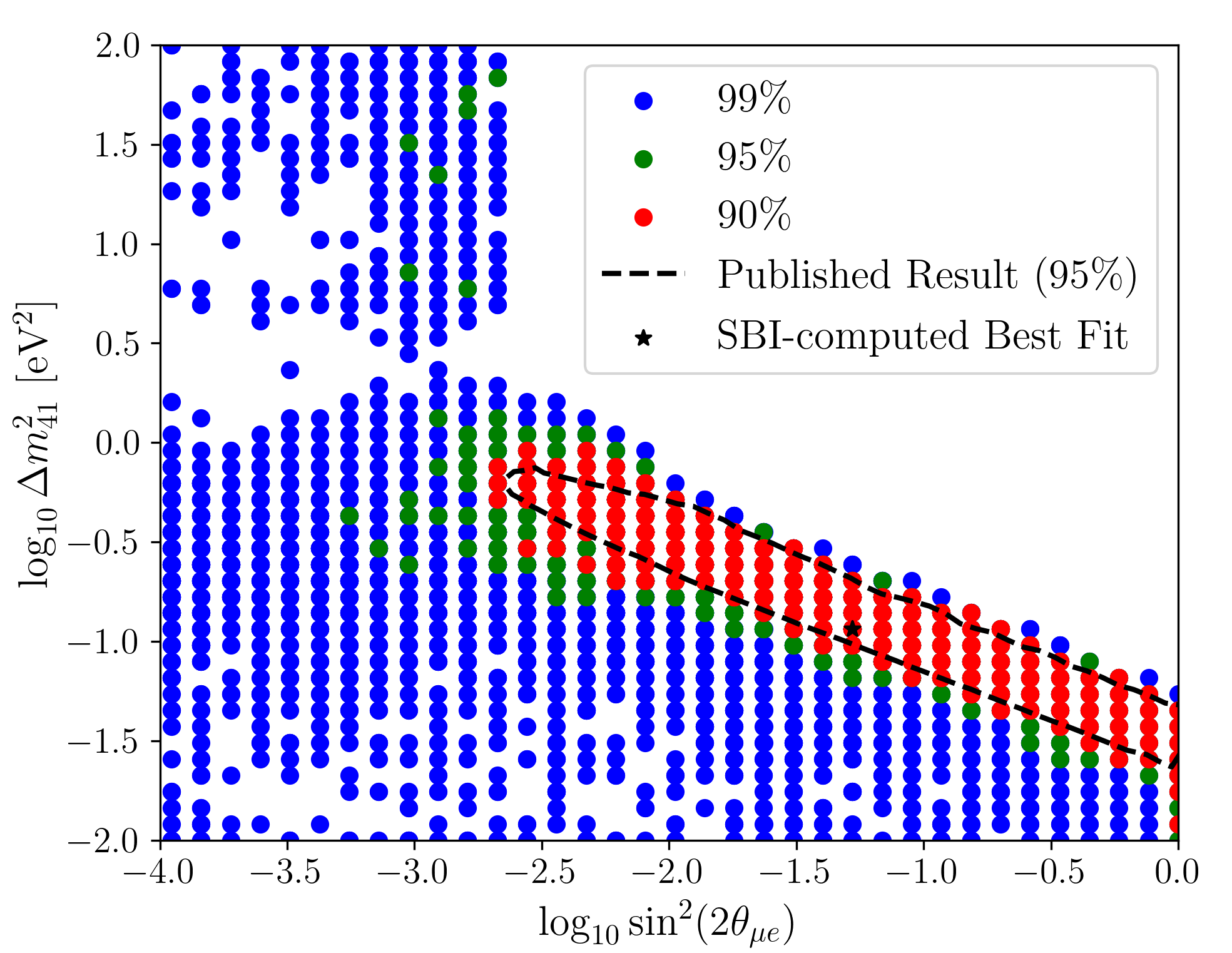}
    
    \caption{SBI-computed joint fit results for MiniBooNE and MicroBooNE, with the $95\%$ CL from \cite{MiniBooNE:2022emn} overlaid for comparison.}
    \label{fig:combined_fits}
\end{figure}

We present joint 3+1 fits to MiniBooNE and MicroBooNE data in Fig. \ref{fig:combined_fits}, using the MicroBooNE systematic uncertainty as published, without correction for the $\nu_\mu$ excess.  A combined fit finds preference for a 3+1 signal, with a SBI-computed best fit of $|U_{e4}|^2 = 0.316, |U_{\mu4}|^2 = 0.362$, and $\Delta m_{41}^2 = 0.115$ eV$^2$. The combined fit is broadly compatible with null at higher confidence levels, consistent with published Wilks'-based results \cite{MiniBooNE:2022emn}. The holes in the accepted region at $99\%$ CL are also compatible with the shape of the MiniBooNE allowed region at high confidence \cite{MiniBooNE:2020pnu}. Similar to the MiniBooNE-only fit, the SBI method finds a slightly larger allowed region than the traditional fit, and we attribute this discrepancy to systematic behavior in both datasets absent from MC simulation. 

We note that the combined fit finds a similar preference ($\gtrapprox 2 \sigma$) for 3+1 with and without the nuisance parameters, indicating that the combined result is largely driven by the MiniBooNE data. 

\subsection{\label{sec:fitsignificances} Fit Significances}
In Table \ref{tab:significances}, we summarize the significance of each fit that prefers 3+1. We compare the empirically calculated SBI-computed significance to the SBI-computed significance assuming that the test statistic $\Delta \tilde \chi^2$ follows a $\chi^2$ distribution with three degrees of freedom, as given by Wilks' theorem. We also compare to published results that use traditional $\chi^2$ methods. 

\begin{table}
    \centering
    \caption{Statistical significance quoted in $\sigma$ for each dataset, computed using SBI with trial-based evaluation, SBI assuming a $\chi^2_{dof=3}$ distribution provided by Wilks' theorem, and the published $\chi^2$-based results \cite{MiniBooNE:2022emn}. ``Combined (n)" denotes the combined fit with the MicroBooNE dataset including additional nuisance parameters.}
    \label{tab:significances}
    
    \begin{tabular*}{\linewidth}{@{\extracolsep{\fill}} lccc}
        \toprule
        & \multicolumn{3}{c}{\textbf{Significance in $\sigma$ (Method)}} \\
        \cmidrule(lr){2-4}
        \textbf{Dataset(s)} & \textbf{SBI (trials)} & \textbf{SBI (Wilks')} & \textbf{Pub. (Wilks')} \\
        \midrule
        MiniBooNE &  3.6 & 1.8 & 4.6\\
        MicroBooNE & 1.8 & 0.4 & N/R$^*$ \\
        Combined &  2.2 &  0.7 & N/R$^*$\\ 
        Combined (n) &  2.1& 0.7& 3.4\\
        
        \bottomrule
    \end{tabular*}
    \vspace{0.5em}
\begin{minipage}{0.9\linewidth}
\footnotesize
$^{*}$Not reported in Ref.~\cite{MiniBooNE:2022emn}.
\end{minipage}
\end{table}

\par For the fits that prefer 3+1, Table \ref{tab:significances} presents a consistent trend in which the empirical SBI-computed significance exceeds the SBI-computed significance assuming the asymptotic approximation given by Wilks' theorem, indicating that the requirements for Wilks' theorem to hold are not necessarily satisfied in these cases. The most clearly violated condition comes from physical limits on model parameters, such as $\sin^2 2\theta_{\mu e} \geq 0$, which can cause the fit to favor a best-fit point consistent with null \cite{Hardin:2022qdh}. Consequently, the resulting $\Delta \tilde{\chi}^2$ is very small, pushing the SBI-computed $\Delta \tilde{\chi}^2$ distribution leftward of the expected $\chi^2$. We conclude that Wilks' theorem is generally not justified in neutrino oscillation analyses, nor in any framework involving bounded physical parameters \cite{algeri2020searching}.  
\par Furthermore, the choice of 3 degrees of freedom is conservative; we expect the empirical distribution to lie to the left of a $\chi^2$ with 3 degrees of freedom due to the degeneracy in $\nu_e$ appearance and disappearance. One major benefit of the trials-based approach is that it avoids the choice of degrees of freedom, leading to more accurate results. 

\par Finally, in Table \ref{tab:significances} we report  considerably lower signal significances for SBI-evaluated results than published. The published statistic is calculated using traditional $\chi^2$ minimization techniques assuming Wilks' theorem. This trend is consistent with the SBI-generated fits being more conservative than the Wilks'-based fits. As explained, we believe both discrepancies arise primarily from the systematic effects in experiments' datasets that are penalized in the SBI framework, lowering the significance of the result. 

\section{PG Tension Computed with Simulation-Based Inference}

As discussed above, it is insufficient to only show that 3+1 gives a better fit than null to the individual and joint samples.   It is necessary to demonstrate that the 3+1  parameters of the individual sets and the joint fit are in agreement.  In this section, we  evaluate the PG tension using our SBI framework. 

\subsection{\label{sec:tension}Evaluating the PG Tension using Experiment-Supplied Systematics}
\begin{figure}
    \centering
    \includegraphics[width=0.85\linewidth]{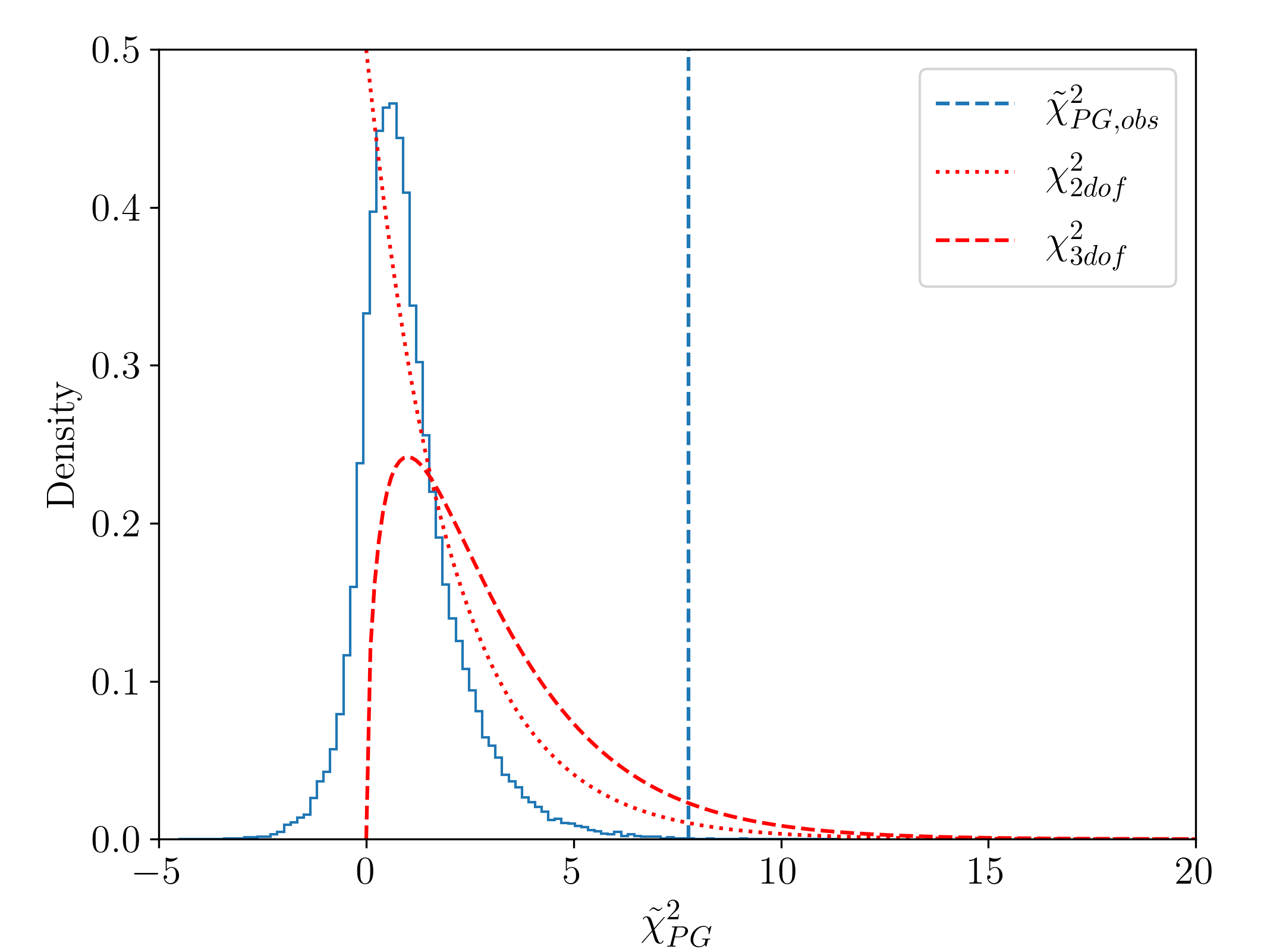}
    \caption{Normalized histogram of $ \tilde{\chi}^2_{PG}$. The blue histogram is constructed from $50,000$ null realizations, and the observed value is shown as the dashed line. Relevant $\chi^2$ distributions are overlaid in red. }
    \label{fig:tension}
\end{figure}

\par Fig. \ref{fig:tension} presents the SBI-approximated $\tilde{\chi}^2_{PG}$ for fits that use the systematic uncertainties provided by each experiment, with no additional pull terms. The distribution roughly lies somewhere between $\chi^2_{dof=2 }$ and $\chi^2_{dof=3 }$, indicating that neither adequately captures the true effective degrees of freedom. As discussed, the distribution is generally shifted to the left of the $\chi^2$ curves, reflecting the impact of physical constraints on the model parameters. Furthermore, a notable feature of the distribution is the presence of negative values. In an exact calculation, $\tilde\chi^2_{PG}$ must be non-negative because the maximum likelihood of the combined fit cannot exceed the sum of the maximum likelihoods of the individual fits. However, in an SBI-based fit, we observe a small amount of negative $\tilde \chi^2_{PG}$. We believe that this arises from approximation error in $\Delta \tilde \chi^2$  due to the machine learning. These negative values are simply an artifact of imperfect network predictions, and verifying the calibration of the trials-based method presented here (see Sec. \ref{sec:calibration}) assuages these concerns.
\par Using the SBI-approximated $\Delta \tilde{\chi}^2$ for each fit to compute $\tilde \chi^2_{PG}$, we empirically calculate a $3.29 \sigma$ PG tension between the MicroBooNE and MiniBooNE datasets. 

\subsection{\label{sec:PGwithpull}Evaluating the PG Tension With MicroBooNE Pull Terms Included}
With the additional pull terms included in the MicroBooNE $\nu_\mu$ samples, we empirically calculate a $2.24 \sigma$ PG tension. This is $\sim1\sigma$ lower than the tension between MiniBooNE and MicroBooNE without the additional pull terms, and comparable to the PG tension measured between MiniBooNE and the MicroBooNE DL datasets (see Appendix \ref{sec:dlanalysis}). The reduction implies that the normalization differences between data and MC in the MicroBooNE $\nu_\mu$ samples act as a systematic difference between the datasets, leading to a larger tension. 

\subsection{\label{sec:discussion}Discussion of Discrepancy between MiniBooNE and MicroBooNE}
A tension exceeding $3\sigma$ between the MiniBooNE and MicroBooNE results indicates a statistically significant disagreement between the two experiments. Such a level of discrepancy suggests that no point in the considered parameter space, which spans both the 3+1 and null hypotheses, provides a consistent description of both the MicroBooNE and MiniBooNE results. This is particularly notable because both experiments operate along the same neutrino beamline and are therefore expected to share similar beam-related and environmental systematic uncertainties.

Moreover, using experiment supplied systematics, we have shown that MiniBooNE and MicroBooNE data independently show deviations from both the null hypothesis as well as the 3+1 model. As a result, the observed tension must arise from unknown effects in one or both experiments, incompatible with one another in their present form.

\par Indeed, after accounting for the disagreement between data and MC in the MicroBooNE $\nu_\mu$ samples, the tension drops to $2.2\sigma$, indicating that a substantial portion of the $>3 \sigma$ tension arose from these deviations. However, some level of disagreement between the experiments remains. 

\section{\label{sec:conclusion}Conclusion}
\par In this work, we developed a novel method for evaluating incompatibility between datasets using SBI under a trials-based approach. The trials-based approach crucially avoids biases from choosing degrees of freedom, a quantity that is often unknown in complex physical situations. Furthermore, the application of SBI speeds up the trial evaluation significantly, permitting substantially larger statistics than is feasible with traditional $\chi^2$ minimization techniques.  

\par To showcase our method, we have explored differences in a 3+1 sterile neutrino fit between MicroBooNE and MiniBooNE data, leveraging machine learning to build a test-statistic. We find that a combined fit to MicroBooNE and MiniBooNE data does not exclude the MiniBooNE allowed region at $\geq 95\%$ CL, showing preference for the 3+1 model when considering the data together. However, a trials-based evaluation of the MiniBooNE and MicroBooNE results using experiment-provided systematic uncertainties presents a $3.3 \sigma$ PG tension between the datasets. Including additional nuisance parameters in the MicroBooNE fit that account for a relative normalization uncertainty between the $\nu_e$ and $\nu_\mu$ samples relaxes the tension to $2.2 \sigma$. 

This case study highlighted an important feature of the SBI method: its ability to suggest the presence of systematic behavior in the data. In this example, the SBI method of constructing tension allows us to quantify the contribution of specific systematic effects to the overall observed tension using a trials-based approach for the first time.  

\par The methods we have demonstrated in this analysis can be directly applied to a full 3+1 global fit to assess the tension between datasets sensitive to neutrino appearance and disappearance, as well as other scenarios in particle physics where multiple experiments exhibit tension with a common theoretical model. These techniques also enable global fits involving more complex sterile neutrino scenarios, which have gained increasing interest given that, as shown here, a minimal 3+1 framework cannot simultaneously describe the data from multiple experiments.

\begin{acknowledgments}
JW, JV, JH, and JC thank MIT for support on this project. AS thanks Texas A\&M University for support on this project. This material is based upon work supported in part by the National Science Foundation Graduate Research Fellowship under Grant No. 2141064.
\end{acknowledgments}

\appendix

\section{\label{sec:dlanalysis} Analysis of MicroBooNE DL Dataset}
The MicroBooNE CCQE sample using Deep Learning (DL) reconstruction is limited by statistics, and, as such, systematic uncertainties do not dominate.  This makes it an interesting comparison to the WC sample. 
In this section, we present joint fit results and tension studies with this dataset for direct comparison with Ref. \cite{MiniBooNE:2022emn}. This sample is sensitive only to $\nu_\mu \rightarrow \nu_e$ and $\nu_e \rightarrow \nu_e$ oscillations, while $\nu_\mu$ disappearance is neglected due to its negligible contribution to the background.  

\begin{figure}[h]
    \centering
    \includegraphics[width=0.85\linewidth]{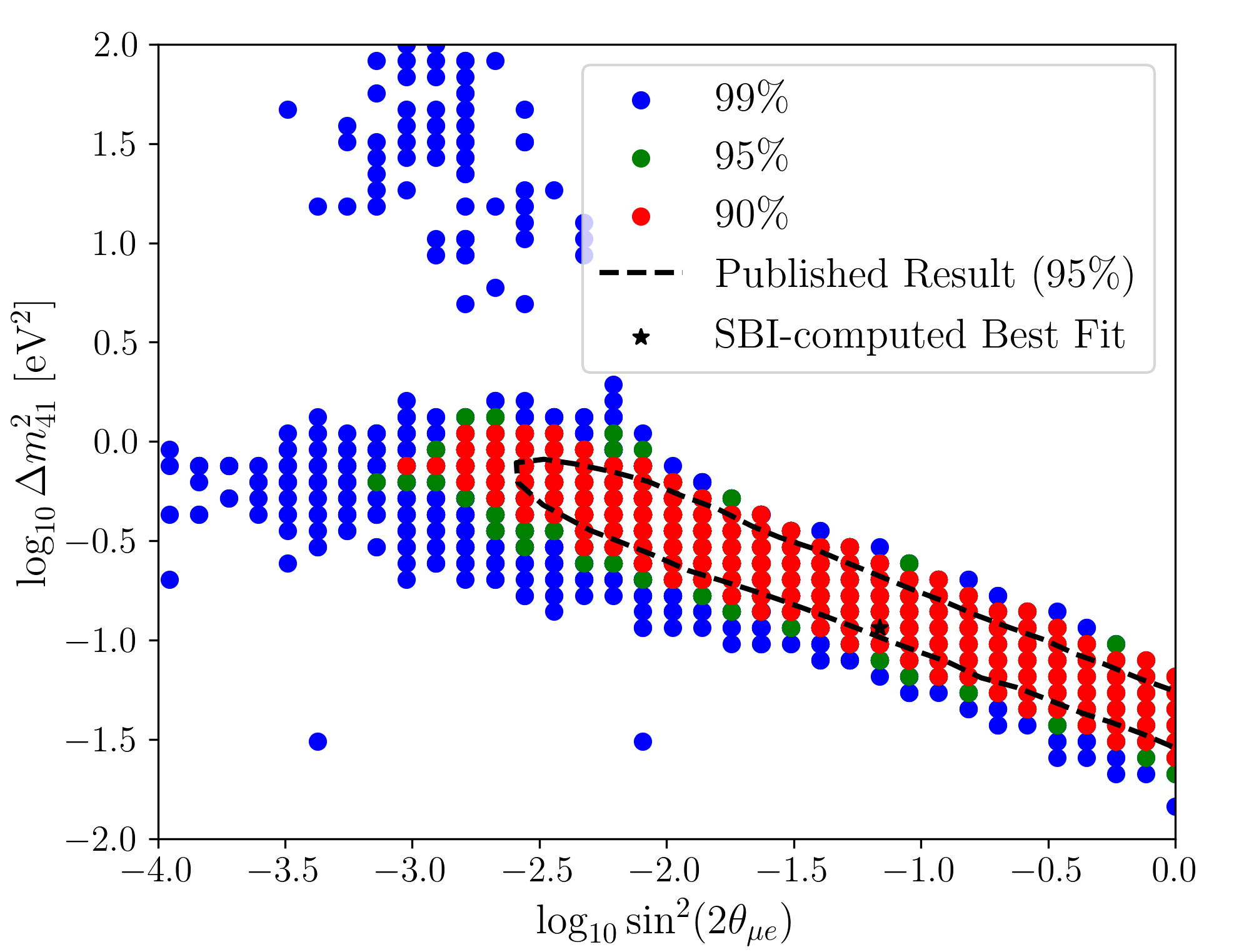}
    \caption{SBI-computed joint fit results for MiniBooNE and MicroBooNE (DL), with the $95\%$ CL from \cite{MiniBooNE:2022emn} overlaid for comparison.}
    \label{fig:DLmBcombined_fit}
\end{figure}

The joint fit to MiniBooNE and MicroBooNE DL data is shown in Fig. \ref{fig:DLmBcombined_fit}. Even at $99\%$ confidence, the fit finds a preference for 3+1, with a best fit of $|U_{e4}|^2 = 0.362, |U_{\mu4}|^2 = 0.362$, and $\Delta m_{41}^2 = 0.115$ eV$^2$. This is a reflection of the low statistical power of the DL dataset; the joint fit is largely dominated by MiniBooNE, which has a strong signal preference. Indeed, the SBI-computed fit significance is $2.9\sigma$, which is almost $1 \sigma$ larger than the fit significance between MiniBooNE and MicroBooNE's inclusive dataset. 

Using SBI-evaluated trials, we empirically calculate the PG tension between the two datasets. We observe a $2.45 \sigma$ PG tension between the MiniBooNE and MicroBooNE (DL) datasets.  

\section{\label{sec:calibration}Calibration of SBI-Evaluated PG tension}
In this section we quantify the calibration of the $\tilde\chi^2_{PG}$ test statistic.

We first note that a Wilks'-based evaluation of the MiniBooNE and MicroBooNE DL datasets using traditional $\chi^2$ minimization techniques yields a PG tension of $2.2\sigma$, comparable to the PG tension observed using a trials-based method within an SBI framework (Sec. \ref{sec:dlanalysis}). This agreement is consistent with proper calibration of the SBI-based method. 

In particular, we verify that the estimated test statistic yields correctly calibrated $p$-values under the null hypothesis (i.e. no sterile neutrino oscillations), ensuring that the quoted significance accurately reflects the probability of observing a given level of tension in the absence of a signal. Under the null hypothesis, fits to a large number of pseudo-experiments should yield $p$-values which are uniformly distributed. A Kolmogorov-Smirnov test comparing the distribution of $p$-values derived from the $\tilde \chi^2_{PG}$ for many null pseudo-experiments to a uniform distribution concludes with a $p$-value of $0.22$, indicating sufficient empirical agreement with a uniform distribution \cite{Massey:1951}. 
\par We further evaluate the statistical power of the PG statistic for both null and alternative hypotheses to confirm that the SBI-based estimate of power is well-behaved. The statistical power of a test indicates the probability of correctly rejecting the null hypothesis. In the context of sterile neutrino oscillations, this is a measure of the model's distinguishing power between statistical fluctuations and genuine physical effects. For $N$ datasets generated under hypothesis $\theta$, the empirical power $\hat \pi$ at significance level $\alpha$ is given by 
\begin{equation}
    \hat \pi(\alpha, \theta) = \frac{n}{N}
\end{equation}
where $n$ is the number of datasets with a $p$-value less than or equal to $\alpha$. $\alpha$ is familiarly known as the Type I error rate, or the probability of falsely rejecting null when the null hypothesis is true. Therefore, for many simulated null realizations generated from $\theta_0$, the power should be equal to the Type I error rate ($\hat \pi(\alpha, \theta_0) = \alpha$). 

\begin{figure}[h]
    \centering
    \includegraphics[width=0.85\linewidth]{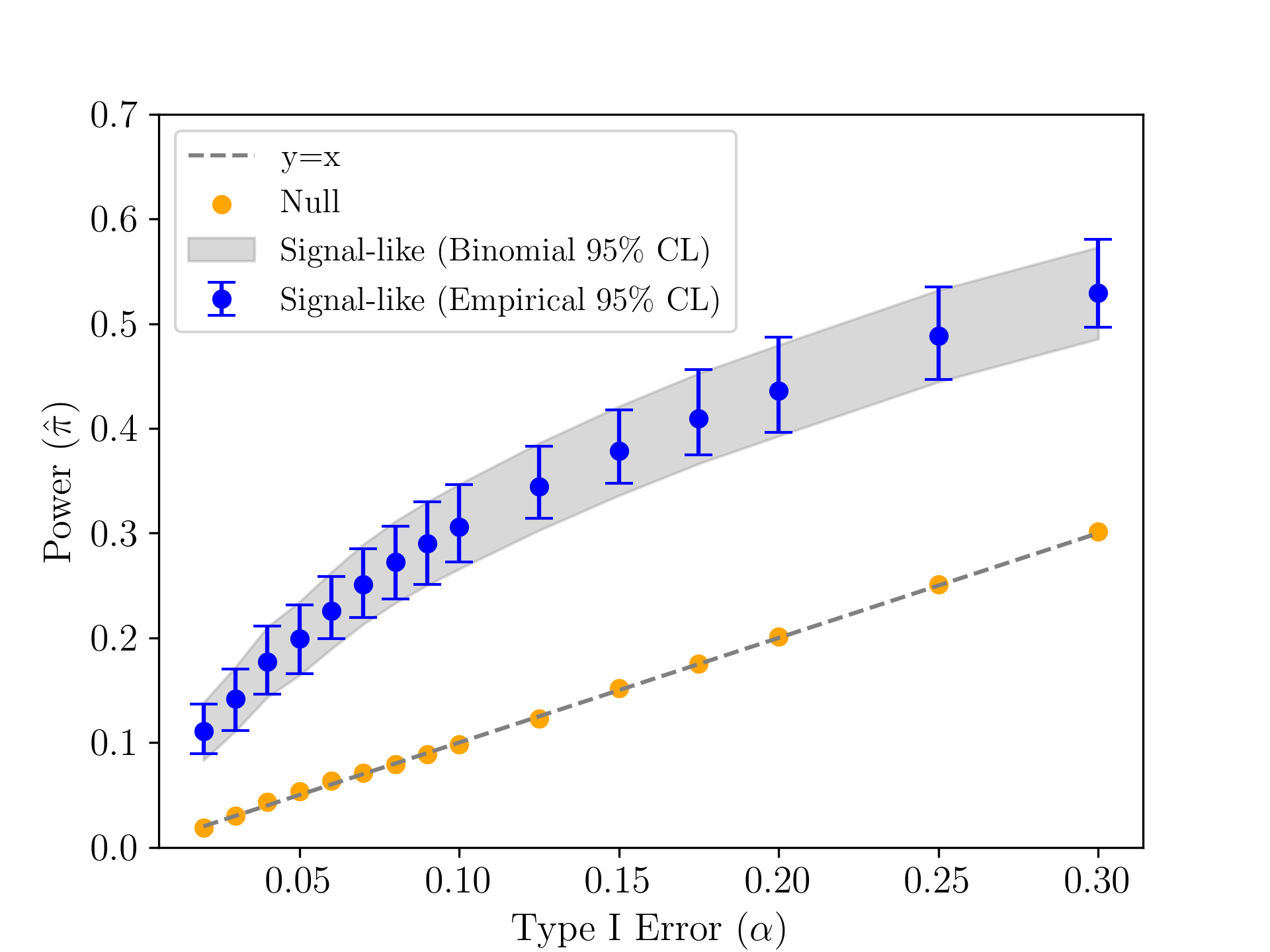}
    
    \caption{Estimated power for various significance levels $\alpha$. The power is calculated for 20 datasets of 500 trials for signal-like realizations (blue), and 1 dataset of 500 trials under the null hypothesis (orange). The signal-like realizations are generated from $U_{e 4}= U_{\mu 4} = 0.168$, $\Delta m_{41}^2 = 0.316$ eV$^2$. The grey band represents the expected $95\%$ CL assuming a binomial distribution, with mean $\mu_\alpha = \text{avg}(\hat \pi_\alpha)$. }
    \label{fig:power}
\end{figure}

Under the alternative hypothesis, the estimated power $\hat \pi$ for $N$ samples should follow a binomial distribution with mean $\pi$ (true power) and variance $\sigma^2 = \frac{\pi(1-\pi)}{N}$ \cite{Cohen:1988}. Without making assumptions about the underlying distribution of the PG tension, the true power of the SBI-evaluated $\tilde \chi^2_{PG}$ is unknown. Instead, we verify that over $M$ different datasets of $N$ samples, the variance of the estimated power $\hat\pi$ at significance $\alpha$ agrees with that of a binomial distribution with $\sigma^2_\alpha = \frac{\hat \mu_\alpha(1-\hat \mu_\alpha)}{N}$, where $\hat \mu_\alpha$ is the empirical mean of the $\hat \pi$ distribution at $\alpha$. We show these results in Figure \ref{fig:power}. We observe excellent agreement between the estimated power for null realizations and expected, and that the measured rejection rates for signal-like realizations exhibit the binomial variance expected for a finite number of trials. 

\section{\label{sec:data-mc}MiniBooNE and MicroBooNE Data}
\begin{figure*}[htb]
    \centering

    \begin{subfigure}{0.48\textwidth}
        \centering
        \includegraphics[width=\linewidth]{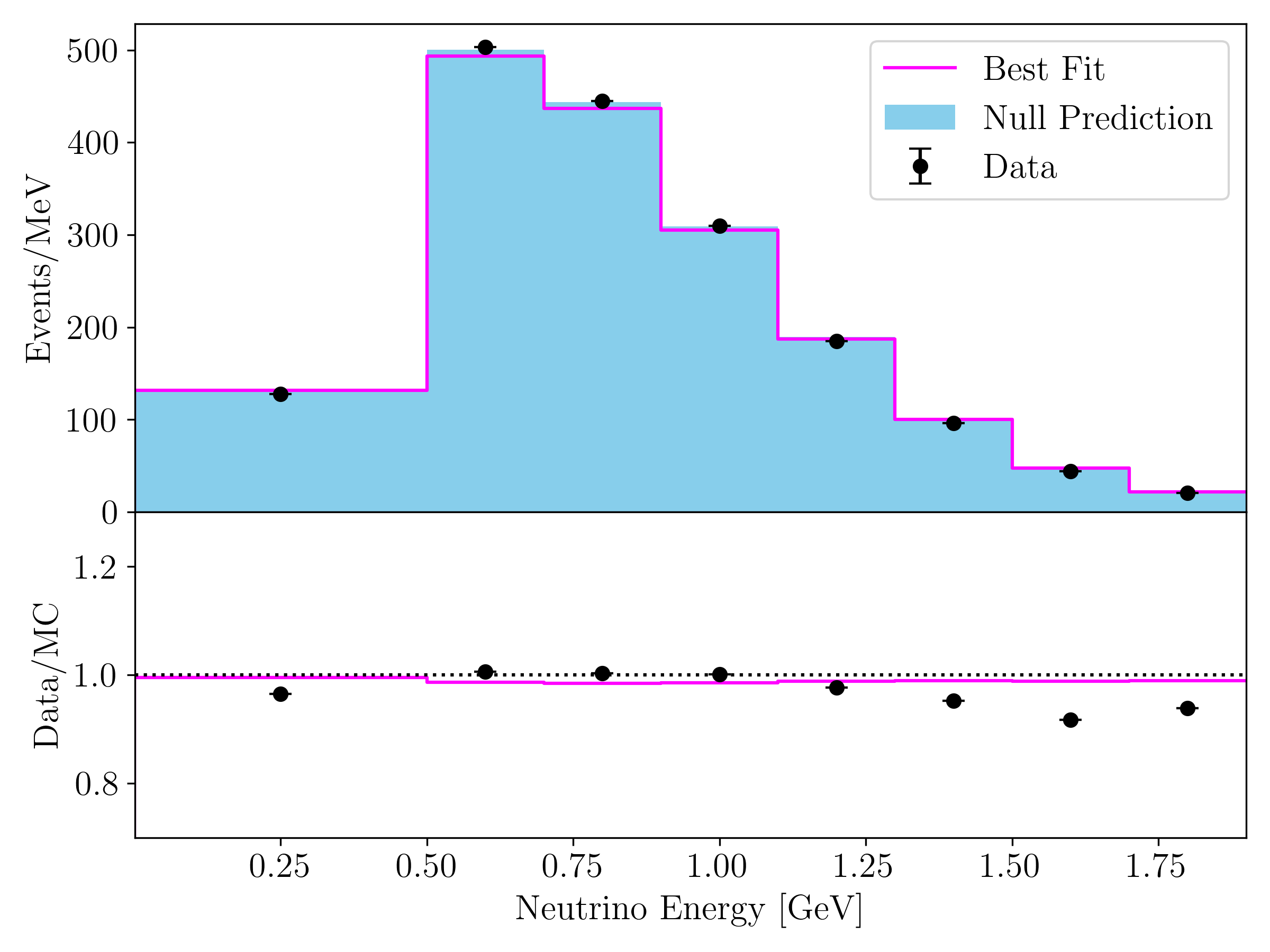}
        
        \caption{MiniBooNE $\nu_\mu$ + $\bar{\nu}_\mu$}
        \label{fig:datanumumB}
    \end{subfigure}
    \hfill
    \begin{subfigure}{0.48\textwidth}
        \centering
        \includegraphics[width=\linewidth]{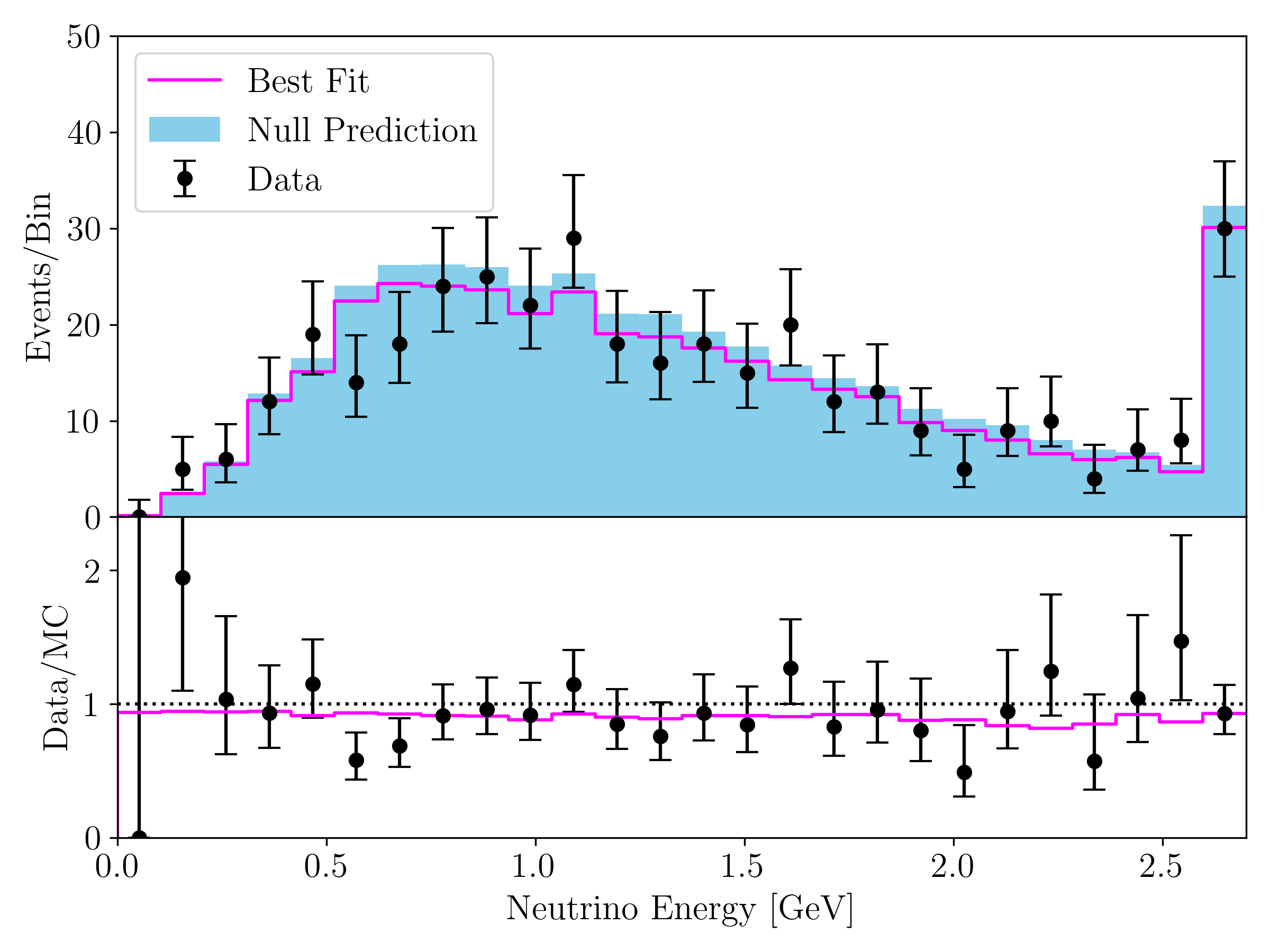}
        
        \caption{MicroBooNE FC $\nu_e$ (WC)}
        \label{fig:nueFC}
    \end{subfigure}
    \hfill
    \begin{subfigure}{0.48\textwidth}
        \centering
        \includegraphics[width=\linewidth]{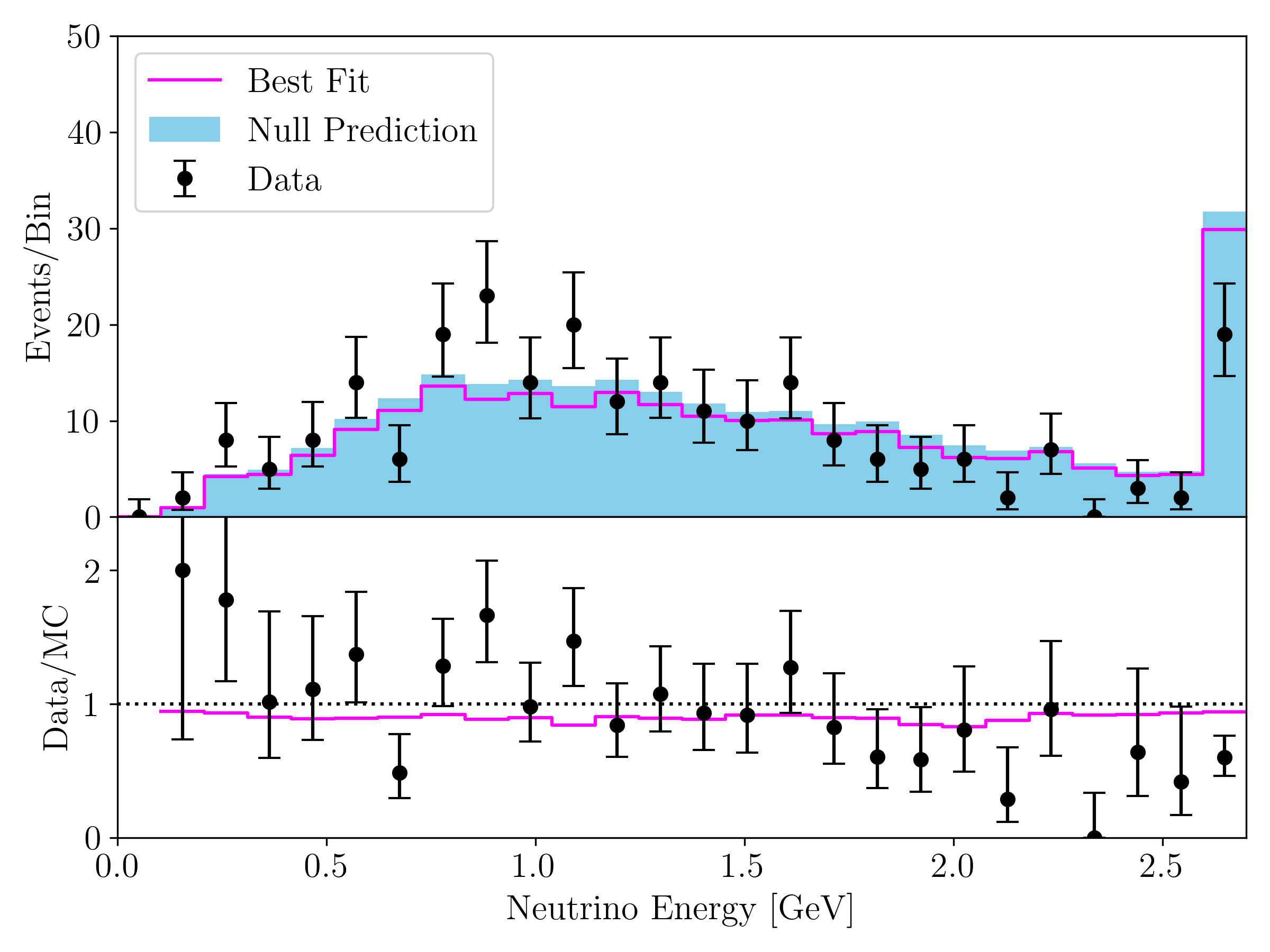}
        
        \caption{MicroBooNE PC $\nu_e$ (WC)}
        \label{fig:nuePC}
    \end{subfigure}
    \hfill
    \begin{subfigure}{0.48\textwidth}
        \centering
        \includegraphics[width=\linewidth]{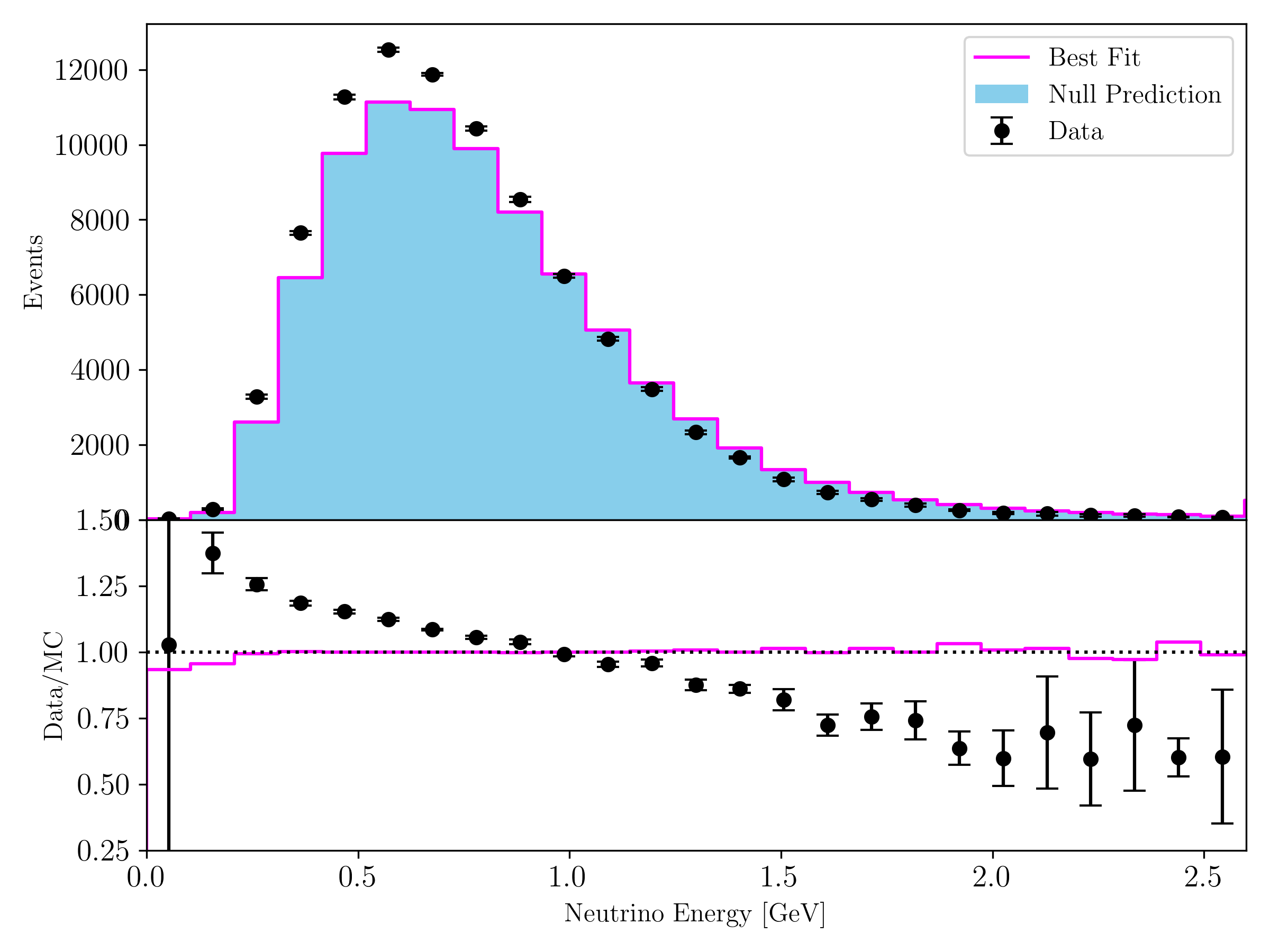}
        
        \caption{MicroBooNE PC $\nu_\mu$ (WC)}
        \label{fig:datanumuPC}
    \end{subfigure}
    \hfill
    \begin{subfigure}{0.48\textwidth}
        \centering
        \includegraphics[width=\linewidth]{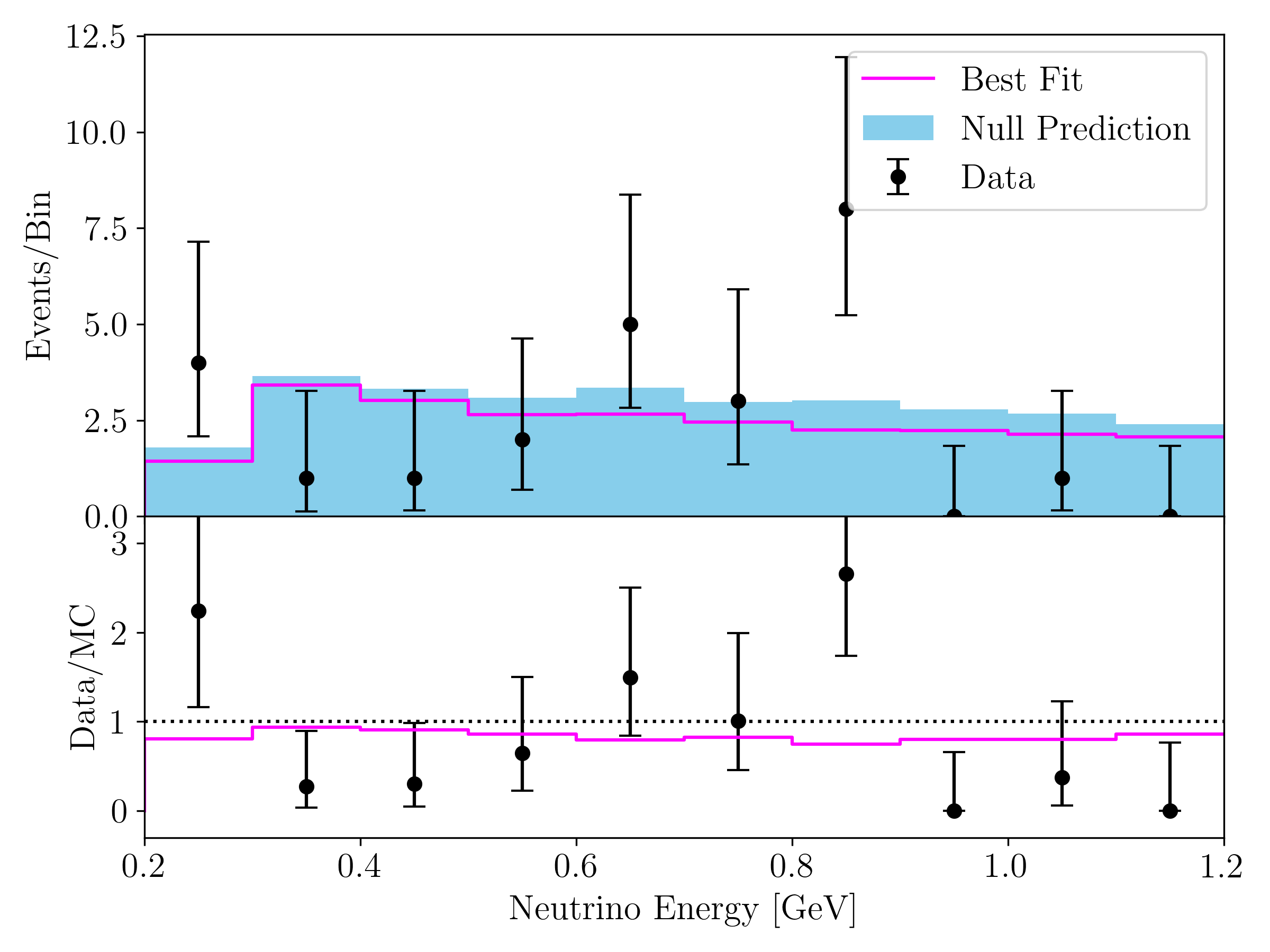}
        
        \caption{MicroBooNE DL}
        \label{fig:dataDL}
    \end{subfigure}
    
    \caption{Data and data/MC ratio distributions for the remaining samples used in this analysis (see Figures \ref{fig:mB} and \ref{fig:WC}), with the Standard Model prediction shown in blue and the SBI-computed best fit 3+1 prediction in pink. Statistical error bars are taken from \cite{MiniBooNE:2022emn}.}
    \label{fig:data_dists}
\end{figure*}

\nocite{*}

\bibliography{bibliography}

\end{document}